\documentclass[journal]{vgtc}                     

\usepackage{latexml}

\graphicspath{{figs/}{figures/}{pictures/}{images/}{./}}

\usepackage{booktabs}
\usepackage{lipsum}
\usepackage{mwe}
\usepackage{ccicons}
\usepackage{mathptmx}
\usepackage{xcolor}

\title{Signals of AI Hallucination: Designing Hallucination-Aware Cues for Embodied Conversational Agents in VR}

\author{%
  Xiaoran Yang, Yang Zhan, Xie He, Yuxuan Huang,
  Yichen Yu, Zhuo Wang, Noboru Matsuda, Qiao Jin
}

\newcommand{\paperabstract}{%
LLM-powered conversational agents (CAs) often present uncertainty and provenance cues alongside their responses to help users assess response reliability and identify potential hallucinations. In immersive environments such as Virtual Reality (VR), CAs often take the form of speech-based embodied conversational agents (ECAs), where uncertainty and provenance cues cannot rely on persistent inline text and may be missed or disrupt comprehension when delivered through speech. We conducted a within-subjects study (N = 24) to compare three designs for presenting the hallucination-awareness information (uncertainty and provenance) in ECAs in VR against a no-cue baseline: embodied cues using gestures and posture, icon cues using visual indicators, and text cues using color-coded text with inline citations. We evaluated how these designs affect users' ability to identify hallucination-related information, trust in the ECA, and interaction experience (immersion and task load). Our results show that all three designs support users in identifying hallucinations. Embodied cues were associated with higher trust and immersion, text cues offered clearer interpretability, and icon cues preserved relatively good interpretability while causing less disruption to immersion compared with embodied cues and text cues. This work contributes to the VR and AI research community by comparing different designs of hallucination cues in immersive ECA settings and examining how they affect users' ability and experiences to identify hallucinations. It also offers practical insights and design implications for developing future hallucination-awareness interfaces for ECA.
}

\newcommand{\paperkeywords}{%
Hallucination Awareness, Embodied Conversational Agent,
User Experience, Human--AI Collaboration, Virtual Reality%
}

\newcommand{\teasercaption}{%
We designed three hallucination-aware cues:
(a) embodied cues (embodied actions-based signals on the embodied
conversational agent),
(b) icon cues (status and source icons), and
(c) text cues (highlighted spans and short labels).%
}

\iflatexml

\makeatletter
\renewcommand{\authorfootertext}{\item[]}
\makeatother

\renewcommand{\manuscriptnotetxt}{}

\else

\onlineid{0}

\vgtccategory{Research}

\authorfooter{
  \item
    Xiaoran Yang is with North Carolina State University.
    E-mail: xyang49@ncsu.edu

  \item
    Yang Zhan is with North Carolina State University.
    E-mail: yzhan3@ncsu.edu

  \item
    Xie He is with Carnegie Mellon University.
    E-mail: xieh@cs.cmu.edu

  \item
    Yuxuan Huang is with North Carolina State University.
    E-mail: yhuang94@ncsu.edu

  \item
    Yichen Yu is with North Carolina State University.
    E-mail: yyu55@ncsu.edu

  \item
    Zhuo Wang is with Xi'an Jiaotong-Liverpool University.
    E-mail: zhuohci01@gmail.com

  \item
    Noboru Matsuda is with North Carolina State University.
    E-mail: Noboru.Matsuda@ncsu.edu

  \item
    Qiao Jin is with North Carolina State University.
    E-mail: qjin4@ncsu.edu
}

\abstract{%
  \paperabstract
}

\keywords{%
  \paperkeywords
}

\teaser{
  \centering
  \includegraphics[width=1\textwidth]{figures/Group49.pdf}
  \caption{\teasercaption}
  \label{fig:sys}
}

\fi

\begin{document}


\iflatexml


\maketitle


\begin{figure}[ht]
  \centering
  \includegraphics[width=\linewidth]{figures/Group49.pdf}
  \caption{\teasercaption}
  \label{fig:sys}
\end{figure}


\section*{Abstract}

\paperabstract

\par\medskip

\noindent\textbf{Keywords:}
\paperkeywords

\par\medskip


\section{Introduction}


\else


\firstsection{Introduction}
\maketitle

\fi





Large Language Model (LLM)-driven conversational agents (CA) have become deeply integrated into everyday workflows for retrieval, creation, and collaborative decision-making, rapidly adopted as on-demand intelligent partners~\cite{wang2024survey}. However, outputs in LLM responses that are linguistically fluent yet contradict established facts or remain inconsistent with the context (defined as hallucination) continue to threaten users' judgment calibration and trust in the system, especially in tasks that require factual verification and high accuracy~\cite{huang2025survey}. For text-based CAs, researchers have progressively developed ways to externalize quality cues through meta-comments, adding supplementary explanations to responses to convey source references and signals of certainty: sentence-level color highlighting~\cite{GoogleBard2023, TheVergeBardDoubleCheck2023}, source and evidence links~\cite{MicrosoftBingNewAI2023}, and clickable provenance cards~\cite{WorkspaceUpdatesRelatedLinks2024}. Study show that users verify hallucinations in textual content more quickly with these additional cues~\cite{nguyen2025hot}. 


When CAs move from text-based chat to immersive environments such as Virtual Reality (VR), they often appear as embodied conversational agents (ECA), animated avatars that share the user's 3D space and communicate through speech, gaze, and gesture~\cite{cassell2001embodied}. This shift replaces persistent, inspectable text with transient embodied signals and redirects attention from a scrollable chat window to the avatar and verbal communication. 
As attention is not centered on the text, chat-style methods for displaying meta-comments on uncertainty and provenance (e.g., inline highlights and evidence links) lack a clear anchor during communication and are difficult to connect with contextual words. Directly migrating the text-based interfaces and cues into VR can break the immersion~\cite{zenner2018immersive} and increase cognitive burden~\cite{liu24toward}. Also, reading these meta-comments aloud would lengthen turns, break the flow of the answer, and raise task load, so it is rarely actionable when needed. It remains unclear how we can design hallucination-awareness interfaces for ECAs that help users notice and interpret hallucination cues in immersive environments.

In this work, we designed and compared three hallucination-awareness interface conditions for presenting hallucination-related information (i.e., response uncertainty and source provenance) in ECAs in VR: \textit{embodied} cues using avatar gestures and posture, \textit{icon} cues using visual indicators, and \textit{text} cues using color-coded text spans with inline citations. As an exploratory study, we evaluated these three designs against a no-cue baseline in a within-subjects study ($N = 24$) to address the following RQs:
\begin{enumerate}[
    label=\textbf{RQ\arabic*.},
    leftmargin=*
]
    \item How do different interface conditions (i.e., embodied cues, icon cues, text cues, and a no-cue baseline) affect users’ actual and perceived ability to identify hallucination-related information in the ECA in VR?
    \item  How do different interface conditions shape users’ trust in the ECA in VR?
    \item  How do different interface conditions shape users’ interaction experiences with the ECA in VR, such as immersion and task load?
    \item What factors shape users’ preferences for different interface conditions in the ECA in VR?
\end{enumerate}

We collected both qualitative and quantitative data to answer the RQs. Our findings show that all three hallucination-awareness cue conditions improved users’ actual and perceived ability to identify hallucination-related information relative to the no-cue baseline. Embodied cues better aligned with the ECA’s natural communication style and helped preserve trust and immersion, yet they could also be easier to miss during ongoing interaction. Text cues offered the clearest interpretability and lowered task demand by making hallucination-related information more explicit and inspectable, while icon cues provided a middle ground between visibility and disruption, preserving relatively good interpretability while causing less disruption to immersion. Users’ preferences were shaped by metaphor consistency, the balance between cue effectiveness and immersion, and interpretability. Our findings suggest that hallucination-awareness design in immersive ECAs is fundamentally about how hallucination-related information enters the interaction. Embodied cues embed it within the agent’s social behavior, text cues foreground it as explicit and inspectable information, and icon cues support more lightweight awareness. Rather than seeking a single optimal cue modality, future designs should treat cue selection as a decision about the interactional role that hallucination information should play.


\section{Related Work}
\subsection{LLM Hallucination and Hallucination Detection}
\label{related_ai_hallucination}
The phenomenon of hallucination in LLMs broadly refers to outputs that are fluent yet inconsistent with facts, inputs, or task constraints~\cite{huang2025survey,Mitchell2023UnderstandingAI}. Existing work~\cite{huang2025survey} has identified three common forms of hallucination: factual-conflict hallucinations, in which generated content contradicts verifiable real-world facts; input-conflict hallucinations, in which the output is inconsistent with the user-provided input or contextual information, such as introducing information not present in the source document during summarization; and context-conflict hallucinations, in which the model contradicts itself within a single response, with later statements conflicting with earlier ones. 
Based on these characteristics, prior work commonly distinguishes between two types of hallucination detection: factuality detection and faithfulness detection~\cite{huang2025survey}. Factuality-based detection checks whether a model output is consistent with verifiable world knowledge or external evidence, while faithfulness-based detection checks whether the output remains consistent with the input source or provided context. Representative methods include Chain-of-Verification~\cite{dhuliawala_chain--verification_2023} and SelfCheckGPT~\cite{manakul2023selfcheckgpt} for factuality, and QAFactEval~\cite{fabbri2022qafacteval} and QuestEval~\cite{scialom2021questeval} for faithfulness.
In this paper, we draw on these definitions of hallucination and use a faithfulness-based detection approach as the backend for identifying potentially hallucinated content. Building on this, we focus on how hallucination-awareness cues should be designed for immersive ECAs by comparing different cue designs.

\subsection{Hallucination-Awareness Designs for CA}
\label{subsec:2.2}
A key challenge in LLM-based systems is not only whether hallucinated content can be detected computationally, but also whether users can notice it and interpret it appropriately during interaction. Users need support in recognizing when a response may be unreliable and in understanding what evidence supports or challenges it~\cite{tomsett2020rapid}. Prior work has therefore explored hallucination-awareness designs that communicate two broad kinds of information to users: uncertainty, which signals that a response or passage may be unreliable, and provenance, which helps users inspect the source, evidence, or justification behind that judgment~\cite{do2024facilitating}.
One common approach is to communicate \textit{uncertainty} or system confidence. Prior studies show that exposing users to confidence estimates can help calibrate trust and reduce overreliance on incorrect AI outputs~\cite{zhang2020effect,bansal2021does,lai2021towards}. In LLM interfaces, this information is often conveyed through confidence thresholds, warning labels, or in-text markings that draw attention to potentially unreliable passages~\cite{karran2022designing,leiser2023chatgpt}. Other designs provide \textit{provenance}-oriented support by revealing why content may be unreliable, for example, through source-related indicators, evidence views, or explicit verification against external knowledge~\cite{leiser2023chatgpt,heo2025halucheck}. Many recent systems combine both. For example, RELIC uses self-consistency to signal potential unreliability while also letting users inspect supporting and contradicting evidence across multiple responses~\cite{cheng2024relic}, and HaluCheck verifies atomic facts against external sources and visualizes both hallucination likelihood and error details~\cite{heo2025halucheck}.

Although many hallucination-cueing methods have been proposed for LLM-based conversational agents, most are designed for non-immersive interfaces such as text chat and web dashboards. Their cues are typically embedded in text through highlighting, footnotes, inline citations, or sidebar annotations~\cite{leiser2024hill, kim2025fostering}. In immersive conversational settings, however, agents often communicate primarily through speech, gesture, and embodied behavior instead of persistent on-screen text. This makes many text-dependent cueing methods difficult to apply directly, and reintroducing text-heavy annotations may interrupt interaction flow and reduce immersion. 
In response to this gap, we explore how hallucination-awareness cues can be designed for ECA in immersive settings such as VR. In our design, hallucination-related information is operationalized through two signal sources: internal indicators from the model’s own generation behavior and external indicators from supporting source evidence. These are presented to users as \textit{uncertainty} and \textit{provenance} respectively. We compare alternative cue designs to study their effects on users’ performance, trust, and experience.

\subsection{Hallucination Risk in Immersive ECAs}\label{RL3}

An embodied conversational agent (ECA) is a virtual character, often with a human-like appearance, that engages users in face-to-face interaction through both verbal and nonverbal channels, including gestures, facial expressions, gaze, and posture~\cite{cassell2001embodied}. Relative to text-based conversational agents, ECAs change both the form and timing of interaction. Rather than presenting responses as persistent text that users can reread and inspect, ECAs often deliver information through spoken dialogue and embodied behavior. Users therefore need to follow content in real time while also interpreting how it is expressed~\cite{elfleet2024investigating,kopp2005conversational}. These embodied signals do not merely add expressiveness; they also shape users' impressions of confidence, competence, and trustworthiness~\cite{waytz2014mind, yang2025tool}. Prior work shows that gaze, posture, intonation, and gesture can affect users’ trust in an agent, with some behaviors signaling certainty and others signaling hesitation or uncertainty~\cite{rohrer2020beat}. 

These effects become stronger when ECAs are placed in immersive environments such as VR. Compared with screen-based ECAs, immersive ECAs share the user’s 3D space and can create stronger bodily co-presence and social presence~\cite{chuah2012increasing, lee2005social}. Users are more likely to experience the agent as being with them, rather than as a character displayed on a separate screen, which can increase perceived realism and intensify social responses to the agent~\cite{lee2003designing, kyrlitsias2022social}. Prior work further suggests that immersive presentation can increase users’ trust in an agent, although this effect depends on design and task context rather than immersion alone~\cite{yildirim2021immersive, kim2018does}. This difference matters for hallucination awareness because the greater perceived trustworthiness of immersive ECAs may increase users’ willingness to accept questionable content without careful checking. When responses are delivered through fluent speech together with gesture and other embodied cues, hallucinated content may feel more credible or be easier to overlook~\cite{brown2023misinformation, pan2025ellma}. Yet little work has examined how hallucination-awareness cues should be designed for immersive ECAs or how different cue designs affect users’ hallucination detection, trust, and experience.

\section{Hallucination-Aware Cue Designs and Implementation}
\subsection{Overview of ECA Setup in VR}
We built a VR prototype to study how hallucination-aware cues can be presented through an ECA, using a classroom scenario as an example interaction context. The prototype was developed in Unity (2022.3.61f1) and runs on Meta Quest 3. It includes a humanoid GPT-4o-driven ECA and a handheld reference panel that users can raise to inspect source evidence when needed. The ECA was equipped real-time lip-sync and a shared set of basic embodied actions. Following prior ECA work on communication-oriented actions~\cite{yang2025tool}, this shared set of actions included greeting gestures directed at the user, pointing gestures used to indicate direction, gestures for describing size or making comparisons, as well as subtle body movements during conversation. These actions are common in everyday communication~\cite{novack2017gesture,goldin2005hearing} and provide a common conversational baseline so that the agent could speak and behave in a socially natural way~\cite{bickmore2008role}.
All those actions were created from self-recorded videos using Animate3D~\cite{DeepMotionAnimate3D}, then converted and imported into Unity’s animation system. During runtime, GPT-4o generated each response in a structured format containing both spoken text and action placeholders (e.g., “[Wave] Welcome!”). The placeholders were mapped to corresponding avatar animations in Unity, while the remaining text was synthesized into speech using Google Cloud Text-to-Speech API. 

Within this setup, we implemented three hallucination-awareness cue conditions: embodied cues, icon cues, and text cues, together with a no-cue baseline. The three cue conditions were grounded in the same underlying detection output, but differed in how they conveyed uncertainty and provenance to users, whereas the baseline provided no such support. All conditions used the same text-to-speech voice to keep the vocal tone consistent. Detailed hallucination detection method and cue designs are described in the following subsections.


\subsection{Representing Hallucination-Related Information for Cue Design}\label{HD}
To support cue design, we represented hallucination-related information through two types of indicators. Internal indicators were model-derived and used to represent uncertainty. External signals were based on the availability of cited source evidence and used to represent provenance.

For internal indicators, we implemented a backend pipeline that generated sentence-level uncertainty signals indicating how likely each sentence in the ECA’s response was to contain hallucinated content. We adopted a self-consistency-based implementation inspired by SelfCheckGPT~\cite{manakul2023selfcheckgpt} because it does not require external databases and can be applied with black-box API access, making it a feasible backend choice for our prototype. Notably, other detection methods (e.g., Adaptive Token Selection~\cite{niu2025robust}) could also be used in future systems, as long as they provide sentence-level signals that can support cue presentation. 
The core idea behind SelfCheckGPT is that factual content tends to remain semantically consistent across multiple stochastic samples generated from the same prompt, whereas hallucinated content is more likely to vary or even contradict itself across samples~\cite{manakul2023selfcheckgpt}. Following this intuition, we generated one primary response for presentation to the user (temperature = 0.0), and then generated 20 additional sampled responses for consistency checking (temperature = 1.0) using OpenAI API (GPT-4o model). Each sentence in the primary response was compared with the sampled responses to estimate how consistently its content was supported across generations. Sentences with lower cross-sample support were treated as more likely to contain hallucinated content~\cite{ellermann2022identifying}. This process produced an internal sentence-level score (logit-derived in our implementation) that served as the backend signal for cue generation. 

Following prior work showing that uncertainty should be communicated in an interpretable user-facing form~\cite{ellermann2022identifying}, we did not expose raw backend scores directly. Instead, we mapped them to three levels: \textit{Confident}, \textit{Uncertain}, and \textit{Unconfident}. We chose a three-level scheme to preserve an intermediate category without requiring users to interpret a more fine-grained numerical representation. These labels indicate how likely a sentence is to contain hallucinated content, based on how consistently its content is supported across sampled responses. \textit{Confident} indicates stronger support and thus a lower likelihood of hallucination; \textit{Uncertain} indicates mixed or borderline support, suggesting possible hallucination; and \textit{Unconfident} indicates weaker support and thus a higher likelihood of hallucination.
These levels then served as the common basis for the cue designs described in the next section. The mapping between backend scores and labels was initially determined based on repeated runs of the scoring pipeline on our study materials. After establishing this preliminary mapping, we invited a domain expert to review the materials and validate the final score distribution. The expert confirmed that sentences labeled as \textit{Uncertain} or \textit{Unconfident} were appropriate targets for further verification, supporting their use as a reasonable reference set. We mapped $P \in [0.4, 0.8)$ to \textit{Uncertain}, $P < 0.4$ to \textit{Unconfident}, and $P \geq 0.8$ to \textit{Confident}. These thresholds were therefore selected through empirical calibration on our study materials. It is important to note that the exact score-to-level mapping was operationalized for our study context rather than treated as a universal standard. 

For the external indicator, provenance was represented through the availability of cited supporting source evidence for the current response. When a response was accompanied by such evidence, the corresponding source webpage was made accessible through the reference panel for direct user inspection. Unlike internal indicators, the external indicator was not designed to produce another judgment about whether the content was hallucinated.

\subsection{Designs of Hallucination-Aware Cues}\label{design}
We developed three cue designs to signal system-provided hallucination-related information (uncertainty level and provenance): \textit{embodied cues}, \textit{icon cues}, and \textit{text cues}. We also included a no-cue baseline condition. In this baseline, the ECA retained its embodied appearance and basic conversational embodied actions, but no hallucination-aware cues were presented (Figure~\ref{fig:CU}). To ensure comparability, the three cue conditions used the same sentence-level uncertainty labels and source-availability annotations. Access to source content was held constant across all conditions: participants could inspect the corresponding source webpage through the same handheld reference panel.
\begin{figure}[t] 
    \centering 
    \includegraphics[width=1\columnwidth]{figures/Group47_6.pdf}
    \caption{Overview of four interface conditions: embodied cues, icon cues, text cues, and a no-cue baseline.}
    \label{fig:CU}
\end{figure}

\subsubsection{Embodied Cues}
We used embodied cues to communicate internal and external indicators because they align with the natural communication style of ECAs. For internal indicators, drawing on McNeill's metaphorical gestures~\cite{mcneill2019gesture}, we designed a reflective pose for \textit{Uncertain}: the ECA tilts its head about 30 degrees downward to the right, raises its right hand to support the tilted head, and places its left hand below the right hand near the abdomen. For \textit{Unconfident} content, our goal was to communicate a stronger likelihood of hallucination. Based on prior work linking low confidence to more closed and withdrawn body language~\cite{brinol2009body}, we designed a defensive posture: the head drops downward and slowly sways from side to side, while both arms are crossed over the chest. We did not design gestures for \textit{Confident} because current LLMs produce a large amount of content in the confident range~\cite{VectaraLeaderboard2024}, so a dedicated confident gesture would be triggered very frequently. Such frequent feedback could distract users from task content and reduce immersion~\cite{gonzales2025behavioral, fuembodied}. Moreover, if the confident gesture were triggered too frequently, it could dominate or interfere with other communicative gestures, making the agent’s overall behavior inconsistent with what people would reasonably expect from an embodied conversational agent in real-world interactions. We therefore treated the absence of a hallucination-specific gesture as the default presentation for \textit{Confident} content.
For the external indicator, our goal was to indicate that cited source evidence was available. We therefore designed a lightweight \textit{citation} gesture to direct users' attention to provenance information. In this gesture, both arms extend forward with the elbows bent at 90 degrees, while the index and middle fingers curl slightly downward and repeat the motion three times. 

When a hallucination-specific gesture was triggered, it temporarily took precedence over the ECA's default conversational actions so that the cue would remain salient. We note that the interpretation of gesture may vary across cultural backgrounds; we return to this point in the limitations section.

\subsubsection{Icon Cues}
Graphical elements have been widely used across many application contexts~\cite{saket2017evaluating, saket2019investigating}, including VR environments~\cite{angus1995embedding}. One of their main strengths is their immediacy and visual clarity~\cite{loffler2014happy}. Building on this advantage, we designed a hallucination cue interface based on graphical elements, using a lightbulb and icons to signal hallucinated content. As shown in Figure~\ref{fig:CU}, the color of the lightbulb was mapped to the confidence level. We adopted a traffic-light color scheme~\cite{tak2014color}, where green, yellow, and red correspond to the three confidence labels: confident, uncertain, and unconfident, respectively. For sentences accompanied by external sources, we directly displayed a website icon, since it is often one of the most recognizable visual markers of a webpage~\cite{cheng2007iconic}. These salient icons were placed next to the lightbulb to make the source of the cited content immediately visible and help users make quick judgments.

\subsubsection{Text Cues}
To understand how hallucination cue interfaces commonly used in text-based conversational agents affect user experience and preference in immersive settings, we drew on Gemini’s meta-commentary presentation style~\cite{GoogleBard2023} as well as common hallucination cueing methods used in text-based CAs~\cite{leiser2024hill}. We converted the ECA’s speech into on-screen subtitles, embedded color-coded background highlights directly within the subtitles, synchronized them with the spoken sentences, and displayed abbreviated website names for sentences supported by external sources. For the background highlights, we again followed prior work on the use of color to convey uncertainty~\cite{tak2014color} and adopted a traffic-light color scheme to represent different confidence levels: green for confident, yellow for uncertain, and red for unconfident. For sentences with external citations, we took inspiration from the way GPT presents web search results and designed adaptive rectangular labels to make the source information more salient, allowing users to quickly identify where the information came from.




\section{Methods}
To answer our RQs, we conducted a within-subjects study comparing the three designs introduced in Section~\ref{design}, together with a baseline condition that did not include any hallucination-awareness cues. 

\subsection{Participants and Setting}
Our study has been approved by the University's IRB. Participants were recruited through university email promotion and offline poster distribution. Each participant who completed the study received a \$20 gift card. 24 participants (12 female, 12 male), aged between 19 and 35 years (M = 23, SD = 3.91), were included in the final analysis. Among them, eight participants were heavy VR users, 13 participants had used VR no more than three times, and three had never used VR before. For the experiences with AI tools, 19 participants understood the use of AI tools, three participants were experts, two participants had passing knowledgeable. For the level of understanding of AI hallucinations, ten participants self-reported that they understood AI hallucinations and can recognize examples based on prior experiences. Four participants had heard the term but did not know its meaning, five had never heard of AI hallucinations, and five reported only a basic understanding of the concept. Before the main study, each participant completed a 20-item screening pre-test covering all four historical topics (see Section~\ref{HCG}) to confirm limited prior knowledge and to check that prior knowledge did not differ systematically across topics. We discuss the trade-offs of this screening in the Limitations section.

\subsection{Hallucination Identification Task Design}\label{HCG}

We used historical content to construct the ECA’s spoken scripts because historical knowledge places a high premium on factual accuracy~\cite{lazonder2022quotation}. To support the within-subject design without repeating the same material across the four conditions, we prepared four historical topics: Ancient Rome, Ancient Greece, Ancient Egypt, and the Renaissance. For each topic, we generated one script using the same prompt template, varying only the topic itself while keeping the requested length, instructional style, and overall structure constant. The prompt asked the model to produce a short teaching-style script that the ECA could deliver in VR and that included both externally referenced statements and potentially hallucinated statements. We referred to a public hallucination leaderboard~\cite{VectaraLeaderboard2024} and adopted a cross-model average hallucination rate of 15\%, yielding five hallucination-related segments per script. These segments were distributed across each script at randomized positions to reduce memory and position effects. All scripts were approximately 330 words ($\pm$5 words).
We then used the same backend pipeline described in Section~3.2 to assign sentence-level uncertainty labels and source availability to the scripts, and to connect them with hallucination-aware cues.

For the hallucination identification task, in each condition, participants watched the ECA deliver the script as a short lesson in VR. Their goal was to monitor the ongoing explanation and indicate any moment at which they felt the current content should be \textit{further checked}. Participants reported such moments by pressing the trigger button on the controller. In our operationalization, content that warranted further checking corresponded to sentences labeled by the backend as \textit{Uncertain} or \textit{Unconfident}. A successful detection occurred when a participant flagged a sentence assigned either label. This operationalization reflects the goal of the study, which was to examine whether hallucination-awareness cues helped users recognize content that warranted additional verification. Participants were not asked to determine whether a sentence was definitively incorrect. Accordingly, the resulting measure captures cue-supported awareness of the need for further checking rather than final factual judgments or verification outcomes. When source evidence was available for a sentence, the corresponding cited webpage could be accessed through the handheld reference panel. Participants could raise the panel to inspect the webpage or ignore it and continue listening, depending on their own verification strategy. We did not require participants to inspect the cited source. In this way, the task captured both participants’ immediate recognition of potentially unreliable content and their optional use of provenance information during verification.

\subsection{Data Collection and Measures}

\subsubsection{Questionnaires}
\label{Questionnaires}
We used self-report questionnaires to measure how different types of hallucination-aware cues may influence the following metrics to answer RQ1 to RQ3:  

We developed two items to assess participants' \textbf{perceived ability} to identify hallucination-related information during the interaction. The first item measured their perceived ability to identify content that required further checking: \textit{``Q1: I can easily tell whether the information provided by the agent contains uncertain or potentially false content that requires further verification.''} The second item measured their perceived ability to identify hallucination-related cues in the interaction: \textit{``Q2: I can easily identify hallucination-related information, such as the agent's uncertainty and the availability of provenance information.''} Both items were rated on a 5-point Likert scale (1 = strongly disagree, 5 = strongly agree).

\textbf{User's trust} with ECA was measured using \textit{Trust in Automation Scale (TAS)}~\cite{jian2000foundations}, which includes 12 standardized items (five of which are reverse-coded). 
The questionnaire adopted a 5-point Likert scale to assess users' level of trust in and perceived reliability of the artificial intelligence system. For the data processing of this questionnaire, negative items were reverse-scored while positive items retained their original scores. 

To evaluate the \textbf{level of immersion} experienced by users under different conditions, we used the Immersive Experience Questionnaire (IEQ)~\cite{jennett2008measuring}. Immersion is a key factor influencing learning and interaction experiences in VR. For ECA in VR, a higher sense of immersion can enhance users' perception of the agent's social presence and promote more natural communication; therefore, this questionnaire is essential for assessing the effectiveness of the system design. The IEQ consists of 31 items divided into six subscales: Cognitive Involvement, Real World Dissociation, Emotional Involvement, Challenge, Control, and General Immersion. Cognitive Involvement measures the user's level of attention, thinking, and concentration during system use or participation. Real World Dissociation assesses the extent to which users forget about the real world during the experience. Emotional Involvement evaluates whether users are emotionally affected by the content of the system. Challenge measures the balance between the difficulty of the tasks and the user's perceived competence. Control assesses whether users feel capable of controlling the interaction and outcomes within the system. Finally, General Immersion measures the overall level of immersion experienced throughout the session. This questionnaire also adopted a 5-point Likert scale for evaluation, with higher values indicating a greater level of immersion.

Measuring the \textbf{task load} involved in hallucination detection is crucial, as task load plays a key role in maintaining attention and preventing information overload during the task. These measurements provide valuable insights for designing more effective hallucination-interaction interfaces. To assess the impact of different hallucination prompt interfaces on users' task load during hallucination detection, we employed the raw NASA Task Load Index (NASA-TLX)~\cite{hart1988development}. This questionnaire includes six dimensions: mental demand, physical demand, temporal demand, performance, effort, and frustration, and is used to evaluate users' task load. Each dimension is scored on a scale from 0 to 100, where higher values indicate a greater level of perceived task load, with the exception of the ``Performance'' dimension, which is reverse-scored. 

\subsubsection{Logs on Hallucination Identification}

To examine whether different interface conditions helped users notice content that the system marked as requiring further checking, we analyzed participants' button-press logs using a confusion matrix~\cite{stehman1997selecting} to answer RQ1. We compared participants' further-check judgments (\textit{Human Perceived Truth}) with the system-assigned sentence labels (\textit{AI Perceived Truth}) that were used to drive the hallucination-awareness cues. In this analysis, the system labels served as the reference set rather than independently verified factual ground truth. Sentences labeled by the backend as \textit{Uncertain} or \textit{Unconfident} were treated as content requiring further checking, whereas sentences labeled as \textit{Confident} were treated as not requiring further checking. Participants' reports, logged through button presses, reflected their subjective judgments during the interaction.

Based on this comparison, we computed true positives (TP; correctly flagging content labeled as requiring further checking), false positives (FP; flagging content not labeled as requiring further checking), true negatives (TN; correctly leaving such content unflagged), and false negatives (FN; failing to flag content labeled as requiring further checking). From these values, Precision indicates how often flagged content matched the system's further-check labels, Recall indicates how much of the system-labeled further-check content participants identified, F1 score captures the balance between Precision and Recall, and Accuracy reflects the overall proportion of matched judgments. These metrics characterize how closely participants' judgments aligned with the hallucination-awareness information provided by the system across interface conditions.

\subsubsection{Semi-structured Interview}
At the end of the study, participants completed a 30-minute semi-structured interview. The interview was used to gather participants’ experiences with and opinions about the different hallucination-awareness cue designs, to help interpret and triangulate the quantitative findings, and to collect their preferences to answer the RQ4. The interview included three parts. First, we asked about participants’ overall impressions and their prior understanding of AI hallucinations. Second, we asked participants to compare the four conditions and explain how each one shaped their attention, hallucination-checking strategies, trust in the ECA, and overall interaction experience. Third, we asked participants about their preferences across the designs and the factors that shaped those preferences.

\subsection{Study Procedure}
We employed a within-subject design, where 24 participants experienced all four conditions. To control for potential order effects, we adopted a balanced Latin square design to counterbalance presentation order, ensuring that each condition appeared equally often in each position and was preceded by every other condition with equal frequency. The historical topics were presented in a fixed order across the study. This allowed us to keep the content progression consistent across participants while varying only the interface condition order; because the conditions were counterbalanced over the fixed positions, no single condition was tied to a single topic. 

This study lasted approximately two hours. First, participants signed an informed consent form and completed a demographic questionnaire. The researcher spent about five minutes introducing the purpose of the study, the study procedure, and the design of each condition. After the introduction, participants wore the headset to experience the virtual classroom environment and learn to use the controller to check the handheld panel and report hallucination. Before the study began, participants were informed of the meaning of each cue.
 They were encouraged to ask questions whenever they encountered something unclear. 
Depending on participants' prior VR experience, this training session typically took around five minutes. After training, participants were given a 5–10 minute break. Once participants confirmed that they did not experience motion sickness and were ready to proceed, each participant received a counterbalanced sequence of experimental conditions and started the hallucination identification task. Each task lasted for five minutes. The participants' objective across all four conditions was the same: mark all the moments they felt ``needed a further verification". Whenever participants felt that certain content required further verification, they could press the trigger button on the right side of the controller to report it. When determining which content required further verification, participants could rely on the cues, inspect the source panel, or continue the task without conducting further verification. The system logged these data and subsequently compared them with the system-assigned labels to assess the degree of alignment between users’ judgments and the system’s labels. 

After each condition, participants were required to complete all the questionnaires shown in Section~\ref{Questionnaires}. A five-minute break was provided between conditions to prevent motion sickness or fatigue. No participants reported severe symptoms after the breaks, and all of them successfully completed the tasks. After completing all four conditions, participants took part in a semi-structured interview lasting approximately 30 minutes.




\subsection{Data Analysis}
\subsubsection{Quantitative Data Analysis}
Before analyzing the data, we assessed the internal consistency of the items in each questionnaire using Cronbach's $\alpha$~\cite{cronbach1951coefficient}. If the score was below 0.7, we removed items with low correlations until Cronbach's $\alpha$exceeded 0.7, indicating an acceptable level of internal consistency.
We also calculated a pooled reliability coefficient across all conditions to verify the consistency of item difficulty and discriminability.

For the questionnaires, reverse-scored items were first adjusted (new score = 6 - original score). The mean score for each dimension was calculated, and the overall score was computed as the average of all items. We then conducted Shapiro–Wilk tests of normality~\cite{shapiro1965analysis} on all quantitative measures.  
For normally-distributed data, descriptive statistics are reported as means (M) and standard deviations (SD), and used one-way repeated-measures ANOVA~\cite{larson2008analysis} to examine differences across the four conditions (Baseline, IC, EC, and TC). For data violating normality, descriptive statistics were reported as median (Mdn) and Inter-quartile Range (IQR), and we applied the Friedman test~\cite{friedman1937use}. 

When the omnibus test revealed a significant difference ($p <.05$), we conducted post hoc pairwise comparisons to identify where the differences lay. For parametric data, we used paired-samples t-tests for between-condition comparisons and reported Cohen's d as the effect size, with values greater than .8 indicating a large effect. For non-parametric data, we used Wilcoxon signed-rank tests for pairwise comparisons and calculated the rank-biserial correlation as the effect size, with values greater than .5 indicating a large effect. We applied the Holm-Bonferroni correction~\cite{holm1979simple} to all post hoc p-values for controlling the multiple comparisons.

\subsection{Qualitative Analysis}

Following Braun and Clarke's reflexive thematic analysis approach~\cite{braun2006using,braun2019reflecting}, we analyzed the interview data to identify recurring patterns in how participants experienced, interpreted, and compared the different hallucination-awareness cue designs. This analysis was used to contextualize the quantitative results and to examine participants' preferences across conditions. 
All interview recordings were transcribed verbatim using speech-to-text technology and de-identified prior to analysis. Transcript segments were indexed by participant ID, interview section, and timestamp to preserve analytic context. We began with open coding to capture participants' interpretations, reasoning, and reported strategies across the four conditions. The research team reviewed the transcripts and generated concise, action-oriented codes linked to participant IDs to support later cross-condition comparison. The first author then iteratively reviewed and refined the initial codes and developed candidate themes. All authors subsequently discussed, reviewed, and revised the themes to ensure that they were coherent, distinct, and well grounded in the data, and collaboratively finalized the theme names and reporting structure. Finally, all authors worked together to build the reporting structure for the qualitative findings and write up the results.


\section{Results}\label{RES}


\subsection{Quantitative Results}\label{HRF}
We conducted Shapiro–Wilk tests for normality on all data. The results showed that, with the exception of the Frustration subscale of the NASA-TLX, the TAS questionnaire, the Accuracy subscale of the confusion matrix and the Cognitive and Control subscales of the IEQ, all other variables failed to meet the assumption of normality. In addition, we assessed the internal consistency of all questionnaires and refined the IEQ based on the reliability results. After this adjustment, all questionnaires demonstrated acceptable reliability ($\alpha > 0.7$) and were considered suitable for use in the study.
%

\subsubsection{User Ability of Hallucination Identification}
\label{QRES3}
\begin{figure}[t] 
    \centering 
    \includegraphics[width=0.8\columnwidth]{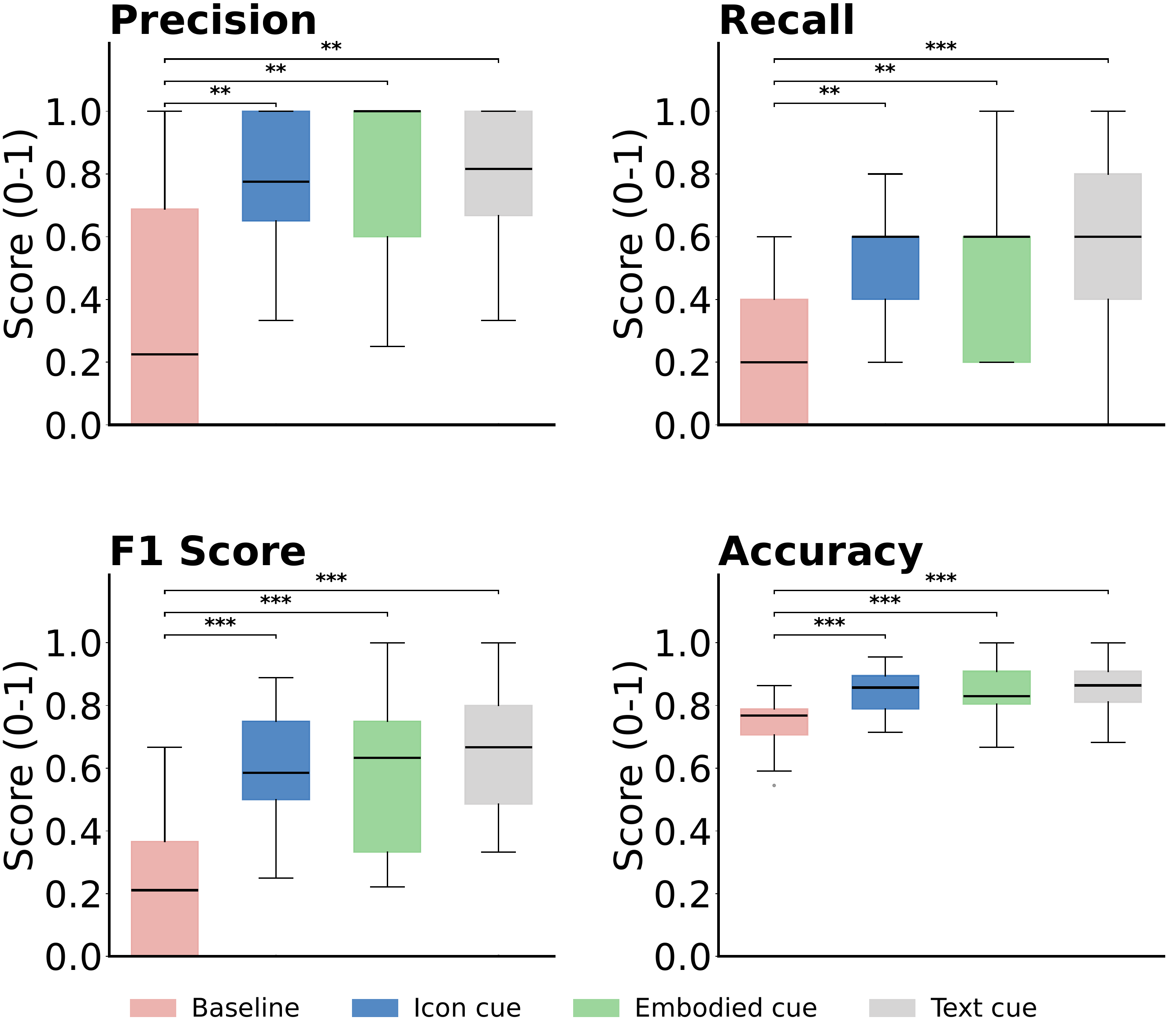}
    \caption{Confusion Matrix Significant Results. Note: Throughout all figures, significance levels are indicated as follows: * $p<.05$, ** $p<.01$, *** $p<.001$.}
    \label{fig:cmb}
\end{figure}

Figure~\ref{fig:cmb} presents the confusion matrix results as users' actual abilities in hallucination identification. The Friedman test revealed significant main effects for Precision ($\chi^2(3) = 21.68$, $p < .001$), Recall ($\chi^2(3) = 32.19$, $p < .001$), and F1 Score ($\chi^2(3) = 35.615$, $p < .001$). A repeated-measures ANOVA also showed a significant main effect for Accuracy ($F(3,69) = 15.65$, $p < .001$).
Post hoc analyses showed that, across all four metrics, each cue condition performed significantly better than the Baseline, while no significant differences were found among the three cue conditions themselves. For Precision, the Baseline condition ($Mdn = 0.23$, $IQR = 0.69$) scored significantly lower than EC ($Mdn = 1$, $IQR = 0.4$, $p <.01$, $r_{rb} = .84$), TC ($Mdn = 0.82$, $IQR = 0.33$, $p <.01$, $r_{rb} = .89$), and IC ($Mdn = 0.78$, $IQR = 0.35$, $p <.01$, $r_{rb} = .90$). For Recall, the Baseline($Mdn = 0.2$, $IQR = 0.4$) condition scored significantly lower than EC ($Mdn = 0.6$, $IQR = 0.4$, $p <.01$, $r_{rb} = .91$), TC ($Mdn = 0.6$, $IQR = 0.4$, $p <.001$, $r_{rb} = 1$), and IC ($Mdn = 0.6$, $IQR = 0.2$, $p <.01$, $r_{rb} = 1$). For F1 Score, the Baseline ($Mdn = 0.21$, $IQR = 0.37$) condition also scored significantly lower than EC ($Mdn = 0.63$, $IQR = 0.42$, $p <.001$, $r_{rb} = .92$), TC ($Mdn = 0.67$, $IQR = 0.31$, $p <.001$, $r_{rb} = 1$), and IC ($Mdn = 0.59$, $IQR = 0.25$, $p <.001$, $r_{rb} = 1$). For Accuracy, the Baseline condition also scored significantly lower than EC ($M = 0.85$, $SD = 0.08$, $p <.001$, $d =1$), TC ($M = 0.86$, $SD = 0.08$, $p <.001$, $d = 1.18$), and IC ($M = 0.84$, $SD = 0.07$, $p <.01$, $d = .88$).

For users' self-perceived ability, figure~\ref{fig:catbtb} shows the results from the subjective questionnaire. Friedman tests indicated significant main effects for both custom items, Q1 and Q2: Q1 ($\chi^2(3) = 22.467$, $p < .001$) and Q2 ($\chi^2(3) = 40.655$, $p < .001$).
For Q1, post hoc comparisons showed that the Baseline condition ($Mdn = 2$, $IQR = 2$) scored significantly lower than EC ($Mdn = 3$, $IQR = 2$, $p = .02$, $r_{rb} = .75$), TC ($Mdn = 4$, $IQR = 1.25$, $p < .01$, $r_{rb} = .95$), and IC ($Mdn = 3$, $IQR = 2$, $p = .02$, $r_{rb} = .86$). TC also scored significantly higher than EC ($p = .02$, $r_{rb} = .73$).
For Q2, post hoc analysis showed that the Baseline condition ($Mdn = 2$, $IQR = 2.25$) scored significantly lower than TC ($Mdn = 4$, $IQR = 2$, $p < .001$, $r_{rb} = 1$), IC ($IQR = 1.25$, $p < .001$, $r_{rb} = .98$), and EC ($Mdn = 4$, $IQR = 2$, $p < .01$, $r_{rb} = .86$).

\subsubsection{User Trust}
\begin{figure}[t] 
    \centering 
    \includegraphics[width=1.0\columnwidth]{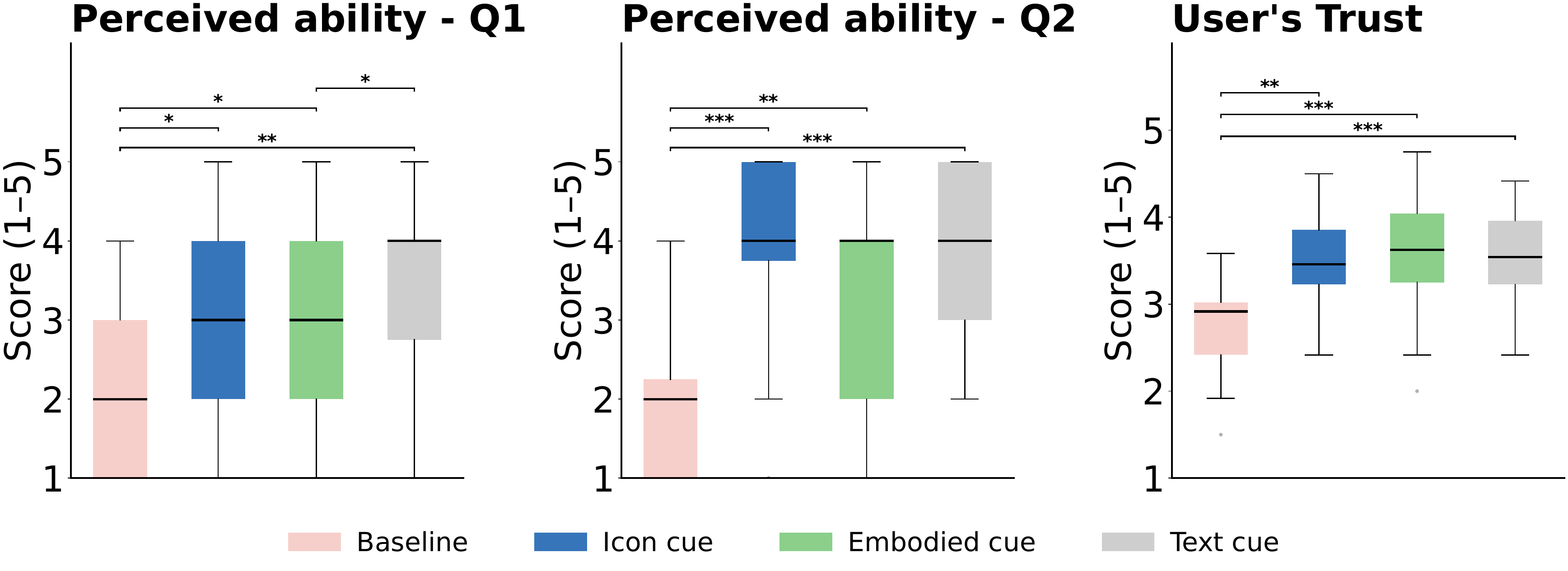}
    \caption{Perceived Ability and User's Trust Questionnaire Results}
    \label{fig:catbtb}
\end{figure}

Figure~\ref{fig:catbtb} presents the results of the Trust in Automation Scale. The repeated-measures ANOVA revealed a significant main effect across conditions ($F(3,69) = 15.46$, $p <.001$). Post hoc analyses showed that the Baseline condition ($M = 2.76$, $SD = 0.48$) scored significantly lower than EC ($M = 3.6$, $SD = 0.65$, $p <.001$, $d = .92$), IC ($M = 3.48$, $SD = 0.54$, $p <.01$, $d = .88$), and TC ($M = 3.55$, $SD = 0.55$, $p <.001$, $d = 1.1$).

\subsubsection{Level of Immersion}\label{QRES2}
\begin{figure}[t] 
    \centering 
    \includegraphics[width=1.0\columnwidth]{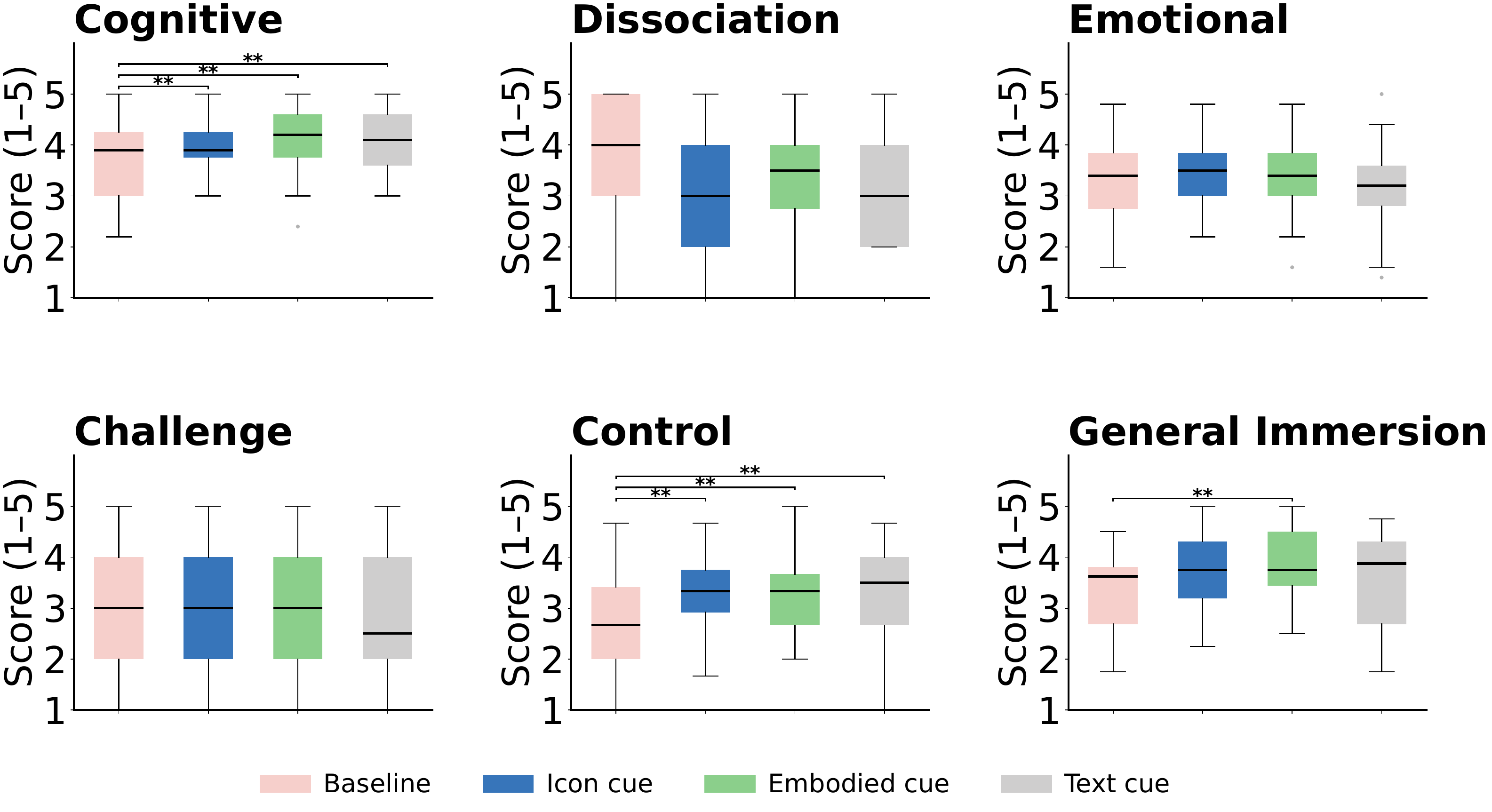}
    \caption{Immersive Experience Questionnaire Results}
    \label{fig:ieq}
\end{figure}

Figure~\ref{fig:ieq} presents the results of the Immersive Experience Questionnaire (IEQ). The repeated-measures ANOVA revealed significant main effects for the Cognitive ($F(3,69) = 4.56$, $p < .01$), Control ($F(3,69) = 6.38$, $p < .001$), and the friedman test revealed a significant effect for General Immersion ($\chi^2(3) = 10.33$, $p = .02$) subscales.
Post hoc analyses showed that, on the Cognitive subscale, the Baseline condition ($M = 3.68$, $SD = 0.76$) scored significantly lower than TC ($M = 4.08$, $SD = 0.63$, $p < .01$, $d = .63$), IC ($M = 4.03$, $SD = 0.54$, $p < .01$, $d = .58$) and EC ($M = 4.1$, $SD = 0.65$, $p < .01$, $d = .60$). On the Control subscale, the Baseline ($M = 2.76$, $SD =1.02$) condition scored significantly lower than EC ($M = 3.25$, $SD = 0.85$, $p < .01$, $d = 0.75$), TC ($M = 3.32$, $SD = 0.88$, $p < .01$, $d = .68$), and IC ($M = 3.27$, $SD = 0.83$, $p < .01$, $d = 0.62$). On the General Immersion subscale, EC ($Mdn = 3.75$, $IQR = 1.06$) scored significantly higher than the Baseline ($Mdn = 3.63$, $IQR = 1.13$), $p <.01$, $r_{rb} =.71$.

\subsubsection{Task Load}

\begin{figure}[t] 
    \centering 
    \includegraphics[width=1.0\columnwidth]{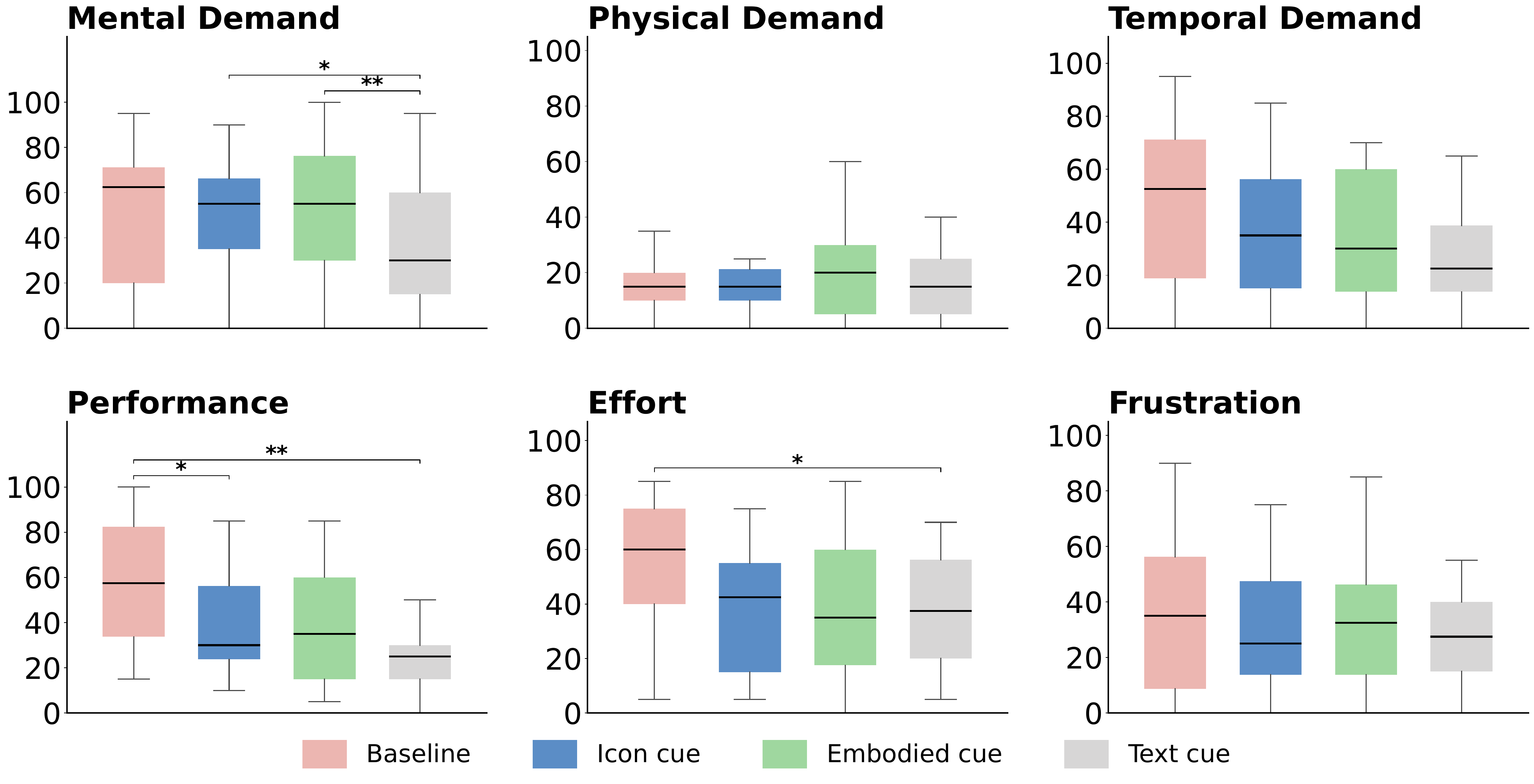}
    \caption{NASA TLX Questionnaire Results(performance subscale is reverse-coded)}
    \label{fig:TLX}
\end{figure}

Figure~\ref{fig:TLX} presents the results of the NASA-TLX questionnaire. The Friedman test revealed significant main effects for the Mental Demand ($\chi^2(3) = 13.43$, $p < 0.01$), Performance ($\chi^2(3) = 17.50$, $p < .001$), and Effort ($\chi^2(3) = 9.29$, $p = .026$) subscales. Post hoc analyses showed that, on the Mental Demand subscale, scores in the TC condition ($Mdn = 30$, $IQR = 45$) were significantly lower than those in the EC ($Mdn = 55$, $IQR = 46.2$, $p < .01$, $r_{rb} = 0.87$) and IC ($Mdn = 55$, $IQR = 31.2$, $p = .0416$, $r_{rb} = 0.61$) conditions. On the Performance subscale, the Baseline ($Mdn = 57.5$, $IQR = 48.8$) condition scored significantly higher than IC ($Mdn = 30$, $IQR = 32.5$, $p = .027$, $r_{rb} = 0.66$) and TC ($Mdn = 25$, $IQR = 15$, $p < .01$, $r_{rb} = 0.86$). On the Effort subscale, the Baseline condition scored significantly higher than TC ($Mdn = 37.5$, $IQR = 36.2$, $p = .033$, $r_{rb} = 0.69$). No significant main effects were found for the remaining subscales, and therefore no post hoc comparisons were conducted for them.

\subsection{Qualitative Results}\label{QRES}

\subsubsection{Theme 1: Familiar cues become intuitive and make instant sense}\label{T1}

In Icon-Cue and Text-Cue conditions, participants appreciated the intuitive awareness of hallucination provided by cues' familiarity, since \textit{``they aligned with the real-world metaphor''} (P23, P24, P26). P27 elaborated on this point, explaining that the color-coded icons were \textit{``very natural because you have been conditioned by the concept of traffic lights since childhood''}, which enabled users to \textit{``distinguish the hallucination quickly even at the first time usage''}. Consequently, many participants indicated that the benefits of familiar cues include efficient detection of hallucinations and ease of understanding (P12, P17, P23). The quick judgment benefited from the intuitiveness further reduced the distraction so that participants could \textit{``focus more on the lecturer avatar''} (P26). 
In contrast, detection efficiency often declined when the meaning of a hallucination cue was unclear and inconsistent with prior common knowledge. In the condition of embodied cues, this inconsistency was caused by overlapping the social meanings and informational cues, which broke the metaphorical understanding and familiarity. For instance, P22 noted that without pre-assigned definitions, both embodied actions simply expressed general uncertainty rather than a specific alert. P5 also explained that \textit{``the prerequisites for understanding gestures were more demanding than for traffic lights, because not everyone might understand that, for example, touching one's head was a sign of hallucination.''} As a result, participants complained that they needed to \textit{``firstly distinguish if the embodied action represented a hallucination''} (P28), which distracted participants' attention from listening to the lecture (P7, P8, P15). 
Additionally, some participants reported a similar perception when comparing the icon-based and text-based citation mark. Icons helped them \textit{``identify if the source website reliable or not at a glance to assist in hallucination detection''} (P24), especially for the familiar websites.


\subsubsection{Theme 2: Embodied cues keep immersive experience but are easy to miss and require further interpretation}\label{T2}

Embodied actions make the avatar feel more like a real conversation partner and preserve immersive experience, but the signal is easy to miss and the meaning can collide with normal social gestures.
Participants generally appreciated that the embodied actions felt realistic, noting that \textit{``the embodied actions were similar to the real-world experience''} (P20, P24) and avoided the use of non-diegetic hallucination cues (P24). This integration improves the social presence of the ECA, as one participant described: \textit{``The interactive experience made me feel more immersive. I felt like I was in a real environment, more like a conversation partner because her movements were similar to the experience of interacting with teachers in real life''} (P26). 
However, with its natural integration, the embodied actions were easily overlooked by participants because they blended in with other \textit{``social gestures while the avatar was speaking''} (P5, P25). To counteract this, a few participants suggested that \textit{``the embodied actions should be exaggerated''} (P11, P26). 
Furthermore, participants desired more informative cues to locate and verify the hallucinations. For example, P28 stated that \textit{``You did not know if the gesture referred specifically to this single sentence, or to this entire block of content and what follows. Therefore, you perceived its localization as being imprecise''}. When comparing the embodied actions with the Text-Cue condition, participants preferred the informative subtitles that helped locate and verify the precise hallucinations. As P27 noted, \textit{``with the subtitles, you could immediately trace back to the exact contents and compared with your prior knowledge framework.''} To further improve localization, some participants suggested that the range of highlighted subtitles could be narrowed to suspicious words and phrases (P6, P24).

\subsubsection{Theme 3: Cue modality and explainable hallucination awareness influences trust on ECA}\label{T3}

When asked what factors affected trust in hallucination cues, some participants stated that the modality of cue could impact their direct perception towards the cues and further the trustworthiness. Some participants perceived the embodied cues as inherently more trustworthy precisely because of their human-like nature. P14 explained that because the embodied cues were \textit{``the most similar to real-world human interaction... it was easier for me to find it acceptable and trustworthy.''} This statement was also supported by P27 and P28, who believed that \textit{``a more realistic appearance would decrease vigilance''} and \textit{``improve its reliability''}. In addition, some participants reported that they would more easily trust in the familiar cue modality (e.g. the signal cues), especially \textit{``If you did not think about it deeply and just scratch the surface, your initial reaction will be to unconditionally trust the AI's prompt''} (P27).
Interestingly, while the embodied modality could increase trust in the system's awareness, A small number of participants expressed the opposite view, suggesting that the embodied modality could reduce their trust—particularly when the embodied signals were used to indicate erroneous content. As P25 noted: \textit{``When the virtual avatar used gestures to signal hallucinations, I did not feel confident learning from it.''}


On the other hand, a key factor influencing participants' trust was the explainability of the hallucination cues. Participants demonstrated greater trust in cues that provided a clear rationale, which convinced them that the system's detection was reasonable and deliberate. As P24 articulated, this clarity was crucial: \textit{``you could clearly understand why the hallucination occurred at this specific point based on its presentation.''} Among the cue conditions, many participants reported that the text cues enhanced their trust in the system's awareness. For example, P20 noted that \textit{``with the subtitles, you could verify the lecturer's statement''}. Similarly, participants' trust increased in the presence of the external source badge. As P28 explained, this feature projected authenticity: \textit{``since it provided a web link and a screenshot, I felt that it was unlikely to be fabricated''}. Other participants echoed this sentiment, believing that the source badge provided an opportunity to conduct secondary confirmation, which thereby improved their confidence.




\section{Discussion}


\subsection{Principal Results}
The quantitative results addressed RQ1--RQ3. For RQ1, the confusion matrix results showed that all three hallucination-awareness interfaces supported better performance in identifying and interpreting hallucination-related information than the baseline, as reflected in higher Recall and Precision. The self-reported items further suggested that the TC condition was associated with greater perceived confidence in identifying hallucination-related information than the EC condition.
For RQ2, our results showed that all three hallucination awareness cues were associated with higher trust ratings than the baseline condition. Although EC received the highest trust ratings among the three cue types, the differences between cue conditions were not statistically significant.
For RQ3, the IEQ results indicated that all cues improved users’ focus on the task and their sense of control over the interaction and the EC condition induced a stronger sense of immersion than the Baseline. The NASA-TLX data showed that the TC condition effectively reduced Mental Demand compared to the EC and IC conditions, and both the TC and IC conditions led to superior task performance over the Baseline.
%
%
The qualitative findings addressed RQ4 and revealed three key factors shaping user preferences: metaphor consistency, the balance between cue effectiveness and immersion, and interpretability. Drawing on both the quantitative and qualitative findings, we derived a set of design implications to inform the development of more effective hallucination cue interfaces for ECAs.

\subsection{Design Implications}

\subsubsection{Balance representational informativeness and visual load in hallucination-awareness design}

In Theme 2, participants described the text cues as helpful because they made hallucination-related information easier to locate and interpret. The NASA-TLX results also showed lower mental demand and effort in the TC condition, suggesting that explicit and inspectable representations can support users' judgments without necessarily increasing subjective burden. At the same time, participants also noted a trade-off: when too many visual elements were presented at once, the interface could feel cluttered and harder to follow, whereas when too little information was provided, it became difficult to determine which content warranted further checking. 
These findings suggest that hallucination-awareness design in immersive ECAs depends less on how much information is shown than on how that information is represented. More explicit and fine-grained representations, such as text highlights and source labels, may better support interpretation and verification, whereas embodied signals may better preserve conversational flow and immersion.
This finding connects to prior work showing that explainability and uncertainty communication depend not only on what information is presented, but also on how it is externalized during interaction~\cite{Wallk2021Review}. From this perspective, designing hallucination-awareness interfaces for immersive ECAs is also about calibrating how hallucination-related information is represented. Future systems could extend our designs by varying the detail, persistence, and explicitness of these representations across contexts and user needs. 

\subsubsection{Highlight the conspicuity of hallucination-related actions to enhance the awareness of embodied cues}
Our results (see Section~\ref{QRES} and ~\ref{QRES3}) suggest that embodied hallucination cues were sometimes difficult for users to interpret because they could be confused with the ECA's ordinary social or conversational actions. This helps explain why participants in the TC condition reported greater confidence in identifying hallucination-related information than those in the EC condition: text-based cues conveyed their meaning more explicitly, whereas embodied cues were more open to interpretation. For example, shaking, head tilting, or subtle hand movements intended to express uncertainty were sometimes interpreted as emotional signals (e.g., disappointment, hesitation, or encouragement) rather than indicators of information reliability. 
These findings suggest that embodied hallucination cues should stay socially integrated while remaining distinguishable from ordinary conversational actions. Rather than avoiding the social action space, designers should make such cues more noticeable without disrupting conversational coherence, for example through clearer motion contrast, more consistent gesture-to-meaning mappings, or brief multimodal reinforcement. At the same time, these mappings may vary across cultural contexts~\cite{kita2009cross}. Designing cues that are both noticeable and contextually interpretable may help users more reliably recognize hallucination-related signals.


\subsubsection{Integrate multimodal hallucination cues to achieve natural fusion and preserve immersion}

The IEQ results (see Section~\ref{QRES2}) suggest that hallucination-awareness cues are less disruptive when they are integrated into the interaction in a natural way. In particular, the EC condition showed higher general immersion than the TC condition, while the broader findings across conditions suggest a trade-off between interpretability and immersion: text cues made hallucination-related information more explicit and inspectable, whereas embodied cues aligned more closely with the ECA's conversational style. In Theme 3, participants likewise noted that cues could undermine immersion when they felt unnatural or overly abrupt. These findings suggest that no single cue type fully resolves the design challenge. Instead, the question is how to combine cue types so that hallucination-related information remains noticeable and interpretable without unnecessarily disrupting the interaction.

These findings suggest the use of a multimodal fusion design, where embodied, visual, and linguistic cues work in temporal, spatial, and semantic coordination. On the embodied level, the ECA could convey uncertainty through subtle facial expressions~\cite{schmidt2024natural} or gaze shifts; on the visual level, the system could display slightly changing confidence indicators or semi-transparent highlights to signal information reliability~\cite{schombs2024robot}; on the linguistic level, hedging expressions (e.g., ``I think it might be…'') could naturally communicate the agent's confidence. Different modalities should complement rather than duplicate one another: embodied signals draw attention, visual cues provide intuitive explanations, and verbal feedback offers reasoning clues. In this way, a multimodal cueing interface can enhance users’ understanding and awareness of hallucinated content while preserving their focus on the ongoing task, thereby reducing the impact of hallucinations on task performance in immersive ECA. 


\section{Limitation \& Future Work}

This study has several limitations. First, our participants were intentionally recruited to have limited prior knowledge of the task content. This helped reduce the confounding effect of topic familiarity and made it easier to examine whether different cue designs could guide users' judgments under comparable conditions. At the same time, this design also means that participants often had little basis for challenging the ECA's responses beyond the hallucination-awareness cues themselves. As a result, the study mainly captures how users respond when system-provided cues are a primary basis for judgment, rather than how users balance those cues against their own domain knowledge. This raises an important concern: cues that are designed to support awareness may also shape, narrow, or even override users' own sensemaking process. In other words, greater cue salience may improve users’ detection of content requiring further checking while also reducing their autonomy in forming independent judgments. Future work should therefore examine how users with different levels of prior knowledge interpret and respond to different cue designs, particularly when they disagree with the system or develop verification strategies that do not rely mainly on interface cues.

Second, we evaluated participants' judgments against the system-assigned uncertainty labels used to drive the cues, rather than against independently verified factual annotations. This choice fit the scope of the study, which focused on whether different cue designs help users notice and interpret the hallucination-awareness information presented by the system. However, it does not establish whether these cues help users identify factually incorrect content when the backend hallucination detection method is imperfect. In real deployment, false alarms and missed detections may distort users' judgments and undermine the reliability of cue-guided verification. Future work should evaluate cue designs against independently established factual reference labels and examine how detection errors influence users' trust and verification behavior.

Third, the study used a tightly controlled, scripted history-learning task. This design allowed us to keep content length, structure, and cue opportunities comparable across conditions, but it also simplified the interaction. Participants mainly listened, monitored the ECA's response, and decided whether the current content required further checking. Many real uses of immersive ECAs are more dynamic: users interrupt, ask follow-up questions, shift goals, return to earlier claims, and interpret information over longer conversations. In such settings, hallucination-awareness cues may interact with memory, turn-taking, conversational repair, and evolving trust in ways that were not captured here. Future work should study these cues in more open-ended and interactive settings, such as tutoring dialogues, collaborative planning, or exploratory information seeking.

Fourth, in the Embodied Cues condition, we did not design a gesture for high-confidence content. This choice was intended to prevent frequently occurring gestures from distracting users and interfering with the presentation of other communicative gestures. Nevertheless, future work could examine whether a selectively presented confidence gesture (e.g., when the estimated confidence level changes or at key points in an explanation) provides useful information without creating excessive movement. Such gestures may affect how users interpret the ECA’s estimated reliability and perceive its personality~\cite{acar2021expressions,neff2010evaluating}. Gesture-validation studies should independently manipulate the presence and frequency of confidence gestures, the system’s estimated confidence, and the factual correctness of the accompanying content. This would help determine whether confidence gestures improve users’ judgment calibration or instead produce habituation, overreliance, or false assurance when confidently presented content is incorrect.

In addition, our empirical findings were obtained in a VR setting and should not be assumed to transfer unchanged to AR or non immersive interfaces. Nevertheless, the broader design trade off identified in our study making hallucination-related information sufficiently noticeable while preserving a fluent interaction experience may inform future research in other media. In AR, one possible direction is to develop socially integrated embodied cues that remain perceptually distinguishable without conflicting with or drawing excessive attention away from the physical environment. For non immersive agents, another possible direction is to develop proactive systems that combine text, icons, voice, and avatar-based signals and adapt cue timing and modality to users’ attentional states and verification needs. These possibilities should be viewed as medium-specific design directions that require further empirical validation.

Finally, this paper examined three cue types as representative designs, but they do not exhaust the design space for hallucination awareness in immersive ECAs. Our prototypes focused on presenting uncertainty and provenance at the moment of response delivery and assumed relatively stable cue meanings. In practice, however, embodied cues may be interpreted differently across cultural and interaction contexts; the same gesture may signal uncertainty in one setting but hesitation or encouragement in another~\cite{kita2009cross}. Future work should therefore move beyond comparing isolated cue types and examine when, for whom, and under what interaction demands different cue representations and combinations best support users’ interpretation.
Despite these limitations, as an exploratory study, we believe this work offers an initial empirical account of how hallucination-awareness cues function in immersive ECAs and provides a valuable starting point for the VR and AI community to design and study more trustworthy AI interactions in immersive environments.

\section{Conclusion}
This study examined how different hallucination-awareness cue designs (embodied cues, icon cues, and text cues, and a baseline condition) in ECAs help users identify and interpret LLM hallucinations, how these designs affect trust and interaction experiences, and what factors shape users’ preferences for hallucination cue design in ECAs. 
The quantitative results showed that all three cue designs effectively improved both users’ accuracy in detecting hallucinations and their trust in the system. Among them, embodied cues strengthened users’ sense of immersion, while text cues improved users’ confidence and efficiency in identifying hallucinations. Icon cues occupied a middle ground, balancing immersion and task efficiency.

The qualitative results showed that embodied cues improved users' trust by strengthening the quality of the human–agent relationship, which in turn indirectly supported hallucination detection. Text cues, by contrast, improved trust and detection performance through higher information density and stronger interpretability. These findings provide an exploratory basis for designing hallucination-awareness interfaces for ECAs.
%



\acknowledgments{
We used OpenAI's GPT-5.2 model solely for language editing. All research content was developed by the authors.}

\bibliographystyle{abbrv-doi-hyperref}

\bibliography{template}

\end{document}